\RequirePackage{fix-cm}
\documentclass[smallextended]{svjour3}       % onecolumn (second format)
\smartqed  % flush right qed marks, e.g. at end of proof
\usepackage{graphicx}
\usepackage{xcolor}
\usepackage{soul}
\begin{document}

\title{Automotive Battery Pack Standards and Design Characteristics: A Review
}


\author{Saeid~Haghbin$^{1}$, \textit{Senior Member,~IEEE,}
 \\
Morteza Rezaei Larijani$^{2,3}$,
 \\
MohammadReza Zolghadri$^{4}$, \textit{Senior Member, IEEE,}
\\
Shahin Hedayati Kia$^{2}$
\\
\\ Corresponding author: Morteza REZAEI LARIJANI, \\ Email: mrl.larijani@gmail.com}


\institute{$^{1}$ AMR Electronique, 45, Allée du petit Plan, Saint-Just, 01250, \email{saeid.haghbin@amrelectronic.com}
%\emph{Present address:} of F. Author  %  if needed
\\
$^{2}$  Laboratory MIS, Université de Picardie "Jules Verne", Amiens, 80039, France. 
\email{mrl.larijani@gmail.com, shdkia@u-picardie.fr}
\\
$^{3}$  Lille Laboratory of Electrical Engineering and Power Electronics, Université de Lille, 59655, France. 
\email{mrl.larijani@gmail.com}
\\
$^{4}$ Department of Electrical Engineering, Sharif University of Technology, Tehran, 14588-89694, Iran. 
\email{and zolghadr@sharif.edu}
 \\
}

\date{Received: date / Accepted: date}

\maketitle

\begin{abstract}
The latest advancements and near-future trends in automotive battery packs, underlying regulatory compliance, and performance requirements are presented in this paper. In response to these specifications, high-level solutions that converge towards a standard architecture for passenger cars are provided. Transition to high-voltage (400V–800V) enables ultra-fast charging above 350 kW, which reduces the charging times to less than 20 minutes. Also, advances in energy density (up to 300 Wh/kg) and battery capacities make advancements in enhancing the electric vehicle's range beyond 1000 km per charge. Key factors such as electrical performance, safety, mechanical integrity, reliability, endurance, environmental conditions, and diagnostics are examined. This study explores the next generation of cost-effective and high-performance battery systems and discovers near-future battery technologies, including sodium-ion chemistry and rare-earth-free alternatives, as well as battery applications in aviation.

\end{abstract}

\section{Article Highlights}
\begin{itemize}
    \item  Review of battery pack for automotive
    \item  Automotive framework development for battery pack
    \item  Providing the latest technological trends and near-future developments for automotive battery pack
\end{itemize}

\section{Introduction}
\label{intro}
\begin{figure}[!t]
\centering
\includegraphics[width=2.5in]{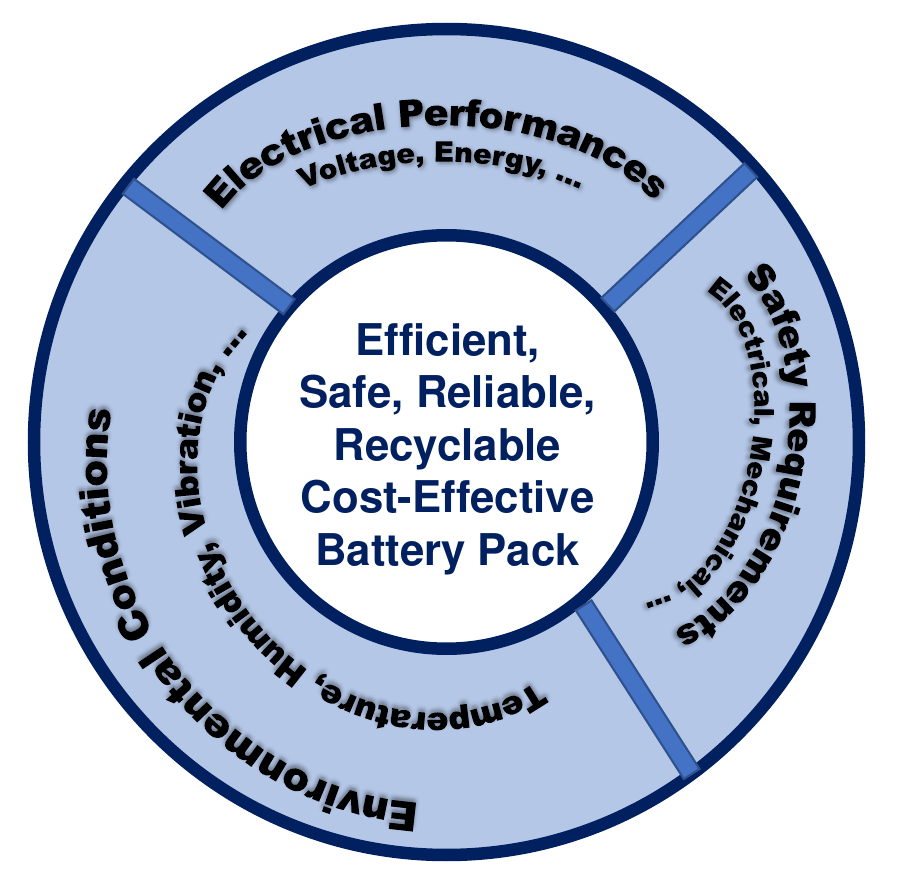}
\caption{Battery pack objectives and constraints.}
\label{Const}
\end{figure}

Battery packs are key components of electric vehicles (EVs) because they operate as the main power supply. Despite recent advancements, further improvements are required to achieve smaller, cheaper, and safer units \mbox{\cite{Matthew,Feng,Naguib}}. \mbox{Fig.~\ref{Const}} shows the ideal battery pack and major constraints. The battery pack, as the main energy storage device for EVs, delivers the required energy and power with a reliable and durable operation that is safe and environmentally friendly \mbox{\cite{Xie,Hu}}. In addition, fast charging is a highly required feature by customers, which adds new aspects to battery pack design, such as busbar temperature monitoring.
Different studies have discussed various aspects of batteries \mbox{\cite{Teichert,Ghaeminezhad,Kuang,Ruffa,Roscher}}.
However, there is still no comprehensive public report showing the overall requirements and development process in the automotive industry. The gap between the automotive industry and academia is well known. This work aims to provide a comprehensive and concise overview and status review of battery packs and near-future trends in the automotive industry. In particular, the required specifications and regulatory standards are more interested.
This review seeks to connect academic research with industry needs by offering a comprehensive overview of automotive battery pack standards, developments in that field, and the essential design considerations. It can serve as a reference for researchers and policymakers in the EV battery technology, which aims to provide more information the latest advancements and identify future challenges that need to be addressed for the development of safer, more efficient, and cost-effective battery systems.

%%%%%%%%%%%%%%%%%%%%%%%%%%%%%%%%%%%%%%%%%%
%%%%%%%%%%%%%%%%%%%%%%%%%%%%%%%%%%%%%%%%%%
Currently, electrification is the subject of heavy investment in the automotive industry. Many activities and developments are underway, but there are limited publicly available data. The main goal of this work is to provide an automotive side view of the topic with a focus on applicable standards from the vehicle level down to the component level. In addition, recent developments and trends are provided at an appropriate level that is suitable in this context. 
This paper is organized into two major parts: different requirements and available solutions that consider regulatory standards. Future trends have been added to each relevant section to ensure greater consistency. Each part is divided into several sections that cover the vehicle level to the component level if required.
The main task of battery storage is to provide energy and power in predefined voltage and current windows. These electrical requirements are explained in Section II. However, there are other specifications, such as the cooling method, charging and discharging capabilities, power quality, electromagnetic compatibility (EMC), and more similar requirements that are addressed in this section.
The following section is dedicated to the battery pack interface with the vehicle. Power and signal connectors and cooling interfaces are the subjects of this section.
The final battery pack should be able to operate in a harsh automotive environment, which is mainly governed by ISO 16750-1 \mbox{\cite{ISO16750-1}, ISO16750-2 \cite{ISO16750-2}, ISO ISO16750-3 \cite{ISO16750-3}}, ISO 16750-4 \cite{ISO16750-4}, and ISO 16750-5 \cite{ISO16750-5}, offering guidance on typical environmental factors that affect electrical and electronic systems in automobiles, along with specified requirements and testing procedures. The battery pack was subjected to extensive environmental testing, such as temperature, vibration, and humidity. This is discussed in Section IV.
Safety is one of the most important requirements of automotive battery packs, as discussed in Section V. The battery pack should be electrically and mechanically safe, and different criteria should be fulfilled as required by the standards. Functional safety is also the main tool for realizing the requirements mentioned.
In response to all the desired requirements, an electrical architecture should be designed and implemented for the battery pack. The battery system architecture for automotive applications converges to a classical structure; however, there have been some slight changes. The battery system structure is discussed in Section VI. The cell technology, battery modules, and power stage of the battery system are presented in this section.
A battery management system (BMS) is a battery system controller that performs measurements, supervision, control, and communication. Section VII describes the BMS and its issues. Some new aspects, such as cybersecurity requirements, were also covered.
Section VIII discusses new applications and general topics. In addition, future studies are presented. The final section presents the conclusion.
%%%%%%%%%%%%%%%%%%%%%%%%%%%%%%%%%%%%%%%%%
%%%%%%%%%%%%%%%%%%%%%%%%%%%%%%%%%%%%%%%%%
\begin{figure}
    \centering
    \includegraphics[width=0.6\linewidth]{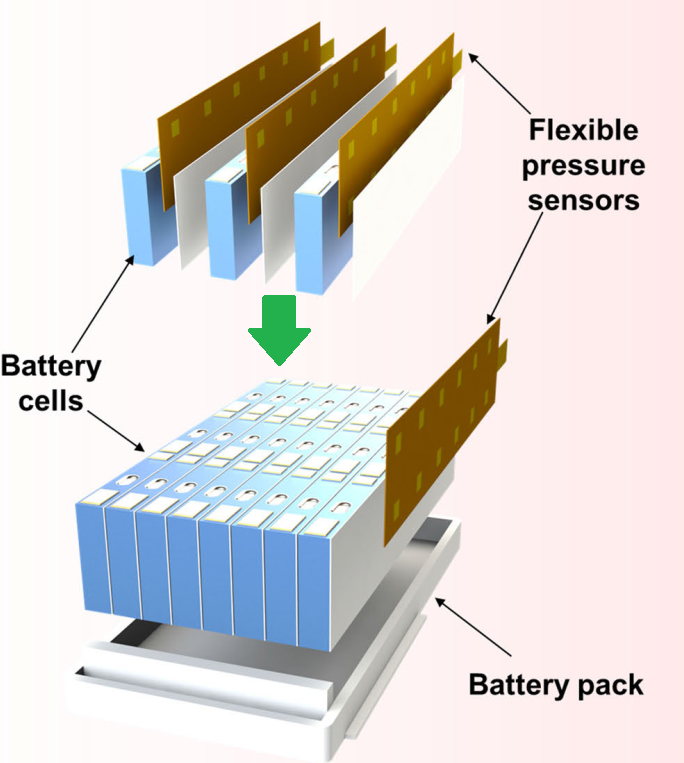}
    \caption{3D view of a battery cell and a battery pack used in an electric vehicle \cite{battery_pack_2024}.}
    \label{battery_pack}
\end{figure}
\begin{figure}
    \centering
    \includegraphics[width=0.5\linewidth]{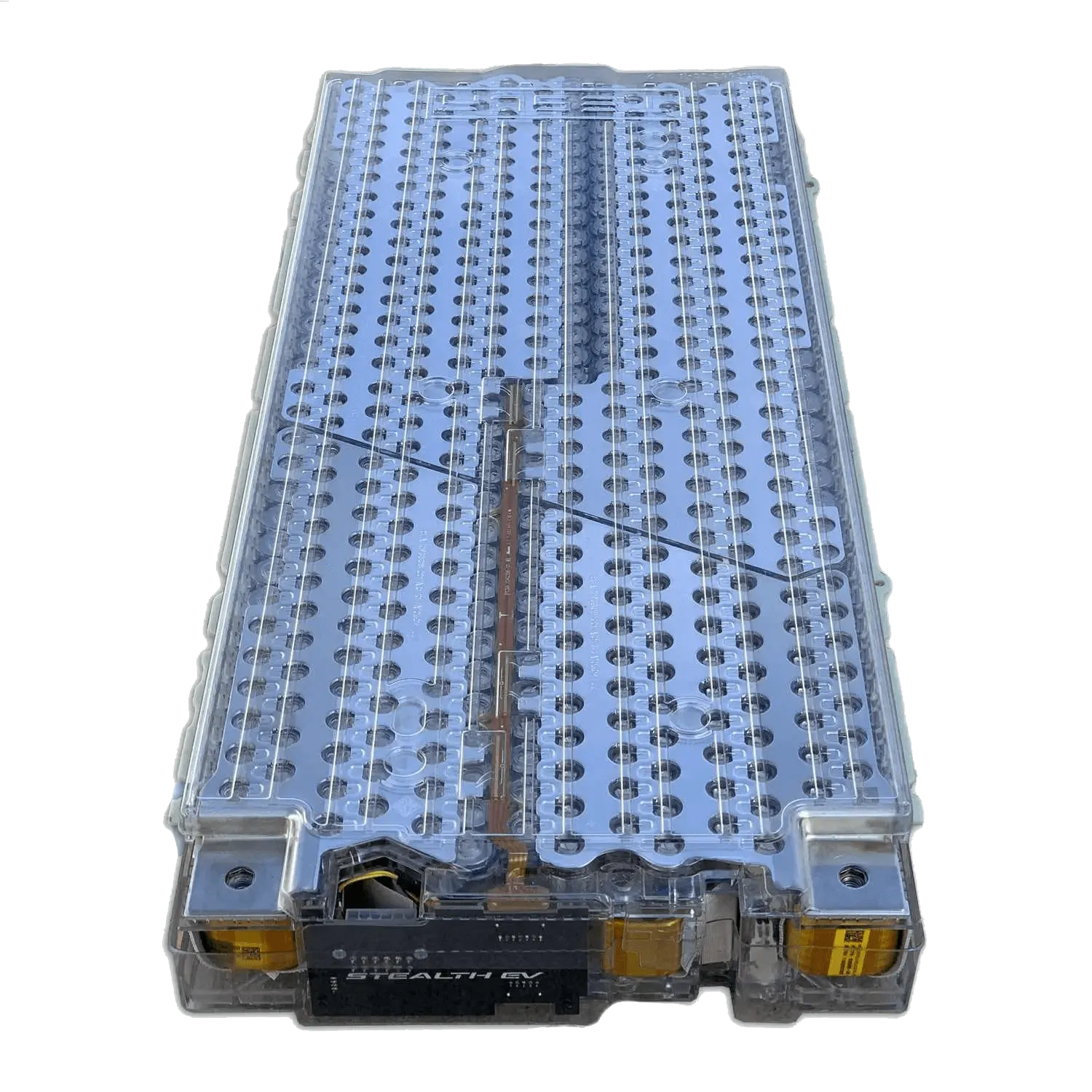}
    \caption{Tesla 5.32~kWh battery module (85~kWh battery pack).}
    \label{Tesla_battery}
\end{figure}
\section{Electrical Requirements}
Fig. \ref{battery_pack} shows 3D view of a battery cell and battery pack which can be used in an EV. Fig. \ref{battery_pack} shows a 3D view of a battery pack in an EV. Fig. \ref{Tesla_battery} shows the Tesla battery module with a capacity of 5.3 kWh, which contributes to a battery pack of 85~kWh. Electrical specifications of the battery pack as the source of traction energy and power are explained in this section. The battery pack should be able to provide the required power and energy for a predetermined lifetime or operational cycle. In addition, the capacity must be within the specified values during different operating conditions and should remain stable during the lifecycle with limited aging factors. All these features are part of the requirements discussed in this section. Electrical characteristics of a battery pack reveal its ability to deliver consistent power and energy throughout its lifespan. The battery system should be stable under different conditions, and consider the minimization of the battery pack aging effects to preserve performance and reliability. These factors play a key role in the efficient and reliable design of the battery systems of EVs.
%%%%%%%%%%%%%%%%%%%%%%%%%%%%%%%%%%%%%%
\subsection{Voltage, Energy and Power}
\noindent A nominal battery voltage of 350~V, 400~V, 700~V, or even higher is used worldwide for  EVs, depending on the vehicle \cite{400_battery}. However, the traction force demand is constantly increasing for different commercial and technical reasons. The price of components reduces over time, and performance is constantly improving in terms of weight and efficiency. Battery systems with higher voltages, that is, 700~V, are expected to become more popular in upcoming years. The vehicle electrical system can be assumed to be a small power generation and distribution system, with similar issues. Increasing the system voltage can reduce losses for a well-designed system. More importantly, with a higher voltage level, it is easier to perform fast charging.\par 
The battery voltage is subjected to a considerable change from the state of being fully charged to the state of depletion. Consequently, nominal values are used for system identification and are not typically used in the component design stage. For example, the minimum and maximum voltage values were used for the design of an electric motor. 
However, increasing the system voltage is not an easy task. There are different aspects, such as safety, that limit the components, both technical and commercial. Hence, it is important to monitor the standard regulations for voltage levels in the automotive industry. ISO 6469 is a widely used safety standard for energy storage systems and protection against failures caused by hazards unique to electrically propelled road vehicle that has four parts ISO6469-1 \mbox{\cite{ISO6469-1}}, ISO 6469-2 \mbox{\cite{ISO6469-2}}, ISO 6469-3 \mbox{\cite{ISO6469-3}}, and ISO 6469-4 \mbox{\cite{ISO6469-4}}. There are two main classes of voltages for automotive applications; classes A and B. However, Class B had more subclasses. Table~\ref{tab:table1} lists automotive voltage classes \cite{ISO6469-3}. As shown in this table, under 60~V of DC is class A, and over 60~V DC to 1500~V is class B. Hence, the battery pack for traction applications is mostly class B. There are different safety requirements for the B1 and B2 classes. Class B2 is targeting 48~V systems.\par
%%%%%%%%%%%%%%%%%%%5
\begin{table}[!t]
\caption{Automotive voltage classes \cite{ISO6469-3}.
\label{tab:table1}}
\centering
\begin{tabular}{|l|l|l|}
\hline
  & \multicolumn{2}{c|}{Maximum operating voltage}\\ 
 %&\\
%\multicolumn{2}{c}{B} &
%\multicolumn{2}{c|}{C} \\
\hline
Voltage class & DC (V) & AC ($V_{rms}$) \\
\hline
A & $0 \leq U \leq 60$ & $0 \leq U \leq 30$ \\
\hline
B & $60 \leq U \leq 1500$ & $30 \leq U \leq 1000$\\
\hline
B1 & $60 \leq U \leq 75$ & $30 \leq U \leq 50$\\
\hline
B2 & $75 \leq U \leq 1500$ & $50 \leq U \leq 1000$\\
\hline
\end{tabular}
\end{table}
%%%%%%%%%%%%%%%%%%%%%%%%%%%%%%%%%%
%%%%%%%%%%%%%%%%%%%%%%%%%%%%%%%%%%%%%%%%%
\begin{figure}
\centering
\includegraphics[width=1.15 in]{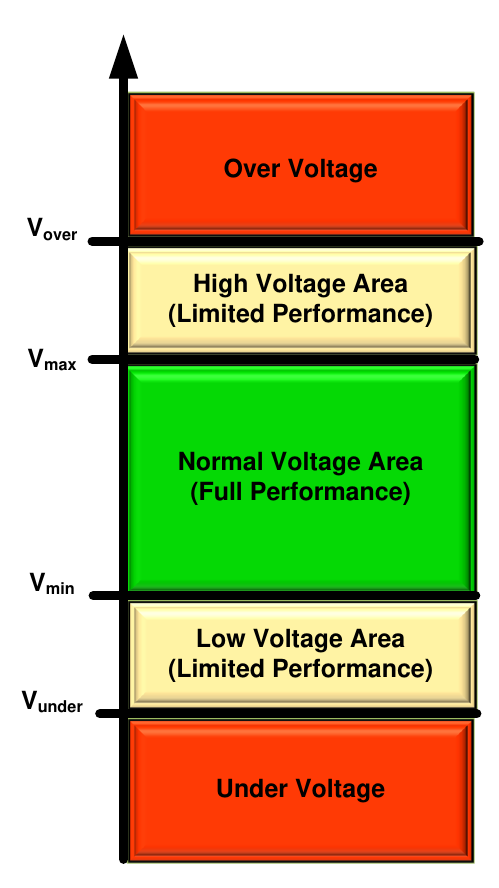}
\caption{Voltage range states of the battery pack.}
\label{Vrange}
\end{figure}
%%%%%%%%%%%%%%%%%%%%%%%%%%%%%%%%%%
In the automotive case, the system status is known for all the cases. In addition, the system's safety status or functional actions should be predetermined and implemented. Hence, the required system performance under different voltage conditions should be clarified, as shown in Fig.~\ref{Vrange} as a typical example. Some areas in the figure may have been integrated. For example, high voltage and overvoltage may occur in the same regions. As a practical example, for a system with a 350~V nominal value, the nominal range can be 300-420~V which the system performs fully in terms of torque and power. The vehicle should be able to operate at low voltages up to 200~V but with limited torque. For high voltages, it can be expected to have a system with reduced performance up to 480~V and afterward, a hard protection action is needed above this voltage level.
Battery energy and power are important electrical requirements that should be specified for the vehicle lifecycle. It is well known that cells are subject to aging, and battery capabilities diminish over the operating cycle. The energy and power of the battery were specified at the beginning and end of its life. There might even be more terms, such as the middle of life, in some companies.
Another important factor is the impact of temperature. Available energy is significantly reduced under cold weather conditions. Additionally, high temperatures pose safety risks. Around room temperatures (15-35~$^oC$), available power and energy are full, while they shall be derated at low and high temperatures \cite{Karlsen,Lajunen1,Lajunen2,Jones,Sundin}.
A maximum power level of 100-200~kW can be assumed for a passenger  EV. Moreover, several ongoing developments are aimed at reaching a 1000~km per charge. It is expected that an energy level of 100-150 kWh can be seen in many upcoming vehicles.
Various standards regulate battery performance, such as the IEC 62660 series, which encourages manufacturers to design interchangeable batteries based on a standardized format, including ISO 62660-1 \mbox{\cite{ISO62660-1}, ISO62660-2 \cite{ISO62660-2}}, ISO62660-3 \mbox{\cite{ISO62660-3}}; ISO 12405 which is designed for the approval of EVs, including ISO 12450-3 \mbox{\cite{ISO3}}, and ISO 12450-4 \mbox{\cite{ISO4}}; UL 2580, which is testing Standard for batteries of EVs, \mbox{\cite{UL2580}}; SAE J2929, which establishes the minimum safety requirements for lithium-ion battery systems used in vehicle propulsion applications connected to a high-voltage powertrain \mbox{\cite{SAEJ2929}}; and GB/Z 1833.1 \mbox{\cite{GBZ1833}}. Further details are provided below.
The powertrain unit of an EV can be a front-wheel drive, rear-wheel drive, or four-wheel drive. In front- or rear-wheel drive there is a single electric motor and inverter, whereas in four-wheel drive, there are usually two electric motors and inverters. Hence, there can be different power demand requirements for the batteries. From a vehicle perspective, each platform is usually designed for three power levels: low, medium, and high duty. For a passenger car, the values for low, medium, and high duties can be 80~kW, 100~kW, and 120~kW peak power, respectively. These numbers were approximate. However, in sports vehicles and new designs, there is a trend for higher power levels. For example, Polestar vehicles may reach a peak power of 500~kW in all-wheel drive configurations. 
The increasing powertrain demand towards higher power levels and easier fast charging are the main motivations for moving towards higher voltage levels like 800~V. However, there is an optimal compromise for component losses and practical safety considerations that limit the maximum voltage.
%%%%%%%%%%%%%%%%%%%%%%%%%%%%%%%%%%
\subsection{Charging, discharging, and life cycle}
\noindent The charging and discharging requirements of the battery pack are directly related to the power demand by the electric motors and the charging time. The battery pack design shall be such that it could meet the required maximum power in traction and regeneration modes. In addition, the charging power is a critical factor for the end users.
A passenger vehicle typically has a lifetime of approximately 15 years or can travel up to 300,000~km \cite{distance},\cite{Americans}. The idea is to design a battery pack to fulfill these requirements. For EVs, based on the end-user usage behavior statistics, these numbers are somehow converted to the number of deep charge and discharge cycles or hours of operation for the electronics. The battery capacity should be sized based on the target mission profile, depending on the type of vehicle, which can be hybrid or electric. The electronics connected to the battery or other components have a lifetime of 40,000-100,000~h. The capacity of a cell is determined based on available statistics, including driver behavior. However, in the specification, the capacity will be specified for extreme conditions, such as low-temperature regions.
Chargers, onboard or offboard, are critical components for the market penetration of EVs. Hence, they have been the subject of extensive research and development \cite{Artex}. Refer to related articles for more details and standards. There are different standards for charging, including auxiliary systemssuch as signaling and earth monitoring, which can be found in references \cite{IEC61851_1,IEC61851_21,IEC61851_22,IEC61851_23,IEC61851_24,IEC62196_1,IEC62196_2,IEC62196_3,SAEJ1772,SAEJ2293_1,SAEJ2293_2,UL2231_1,UL2231_2,UL2251,UL2202,GBT18487_1,GTB1847_2,GTB1847_3,CHAdeMO}.
Chargers are classified as either conductive or wireless. Wireless chargers are under development, but they still require further improvement in terms of cost and weight. Hence, they are not popular. Conductive chargers are 
the dominant solution. However, some attempts have been made to introduce other forms of charging, such as charging via roads, to reduce the required battery capacity.
The onboard chargers are in the power range of 3-22~kW. Because of its cost and weight, it is difficult to reach higher power levels. However, with new power switches, such as SiC devices, there is a good chance to see higher power levels soon. 
Offboard chargers are commercially available in the MW range for many companies. For passenger cars, the trend is towards 350~kW chargers. However, currently, most vehicles are not ready to absorb this power level. There are power limits exist for the battery pack, connectors, and connections. The subject is under extensive development by automakers, and a power level of 200~kW is currently available for several vehicles.
Another class of chargers is integrated motor drive and battery chargers. This technology provides higher power charging available onboard, but has some practical difficulties for implementation. Because they are non-isolated, safety and power quality issues are challenging \cite{Haghbin1}. However, there are some commercially available products like ZOE that, use integrated motor drives and battery chargers. This alternative has a good chance of being widely used shortly because it offers an onboard fast charger with a low price and compact volume. 
The battery pack should be able to operate under predetermined conditions with specified charging and discharging levels in all temperature and environmental conditions. Usually, the lifetime is determined for as the beginning of the lifetime, the middle of the lifetime, and the end of the lifetime. In addition, based on the available internal data, there are company-specific requirements in line with the standard requirements.      
%%%%%%%%%%%%%%%%%%%%%%%%%%%%%%%%%%%%%
\subsection{Power Quality}
\noindent Several components are connected to the battery system, such as the traction inverter(s), DC/DC converter(s), charger, compressor, and heater. This DC distribution system can reach a power level of 500~kW in some vehicles. Hence, it is important to carefully consider power quality and transient issues. In new designs, these topics are gaining more attention in the automotive community, especially at higher voltage levels, such as 800~V or 1200~V.
The power quality or transient issue can be classified into two major groups: limiting the generation of transients or harmonics by components, and the ability to operate under a certain level of transients or harmonics. Different mechanisms or phenomena are associated with power quality or transients. Further details are provided in the following list.
\begin{itemize}
  \item Voltage and current ripple: The ripple over the DC bus or other parts of the system has several consequences, such as increased losses, reduced component lifetime, and EMC issues. The way ripple affects the EMC has a complicated nature, and some activities are ongoing in some automotive companies to correlate the ripple to EMC performance. Hence, it will be more meaningful to set ripple requirements after establishing a more concrete correlation between the ripple and EMC phenomena.
  \item Voltage and current transients: The transients due to low-frequency switching or high-frequency switching have different consequences on the system. Over voltage transients may damage components, especially semiconductor devices. In addition, they have an adverse impact on the EMC performance. Another impact is the stress on insulators in different parts of the system.
\end{itemize}
The battery itself is not a transient or ripple generator. Batteries can absorb transients and ripples to some extent. However, this may lead to issues with the cell lifetime. Extensive research has been conducted on this topic, but it is still not clear how ripples or transients affect cell quality over time \cite{Brand}.
In one case, the battery pack can cause dangerous transients in the system when the battery contactor is suddenly opened as an action in response to a fault like overcurrent. In this case, the electric motors and inverters are forced to an active short circuit state, and the system experiences severe overvoltage. The system design should be such that all components connected to the DC bus can survive this case \cite{Chandran}. 
Excluding the contactor open case, the battery pack was a victim of transients and ripples. Hence, part of the pack requirement is to set a limit on the ripple and transients. As mentioned earlier, this topic is not yet mature, and more work is needed to correlate the ripple and transients with the EMC performance and lifetime of the components. Conservative solutions are typically used in this regard. Naturally, the standard regulations are is not mature. There are few standards addressing topics such as\cite{ISO7637_1,ISO7637_2,ISO7637_3}, but as mentioned, more work or regulations are needed.
%%%%%%%%%%%%%%%%%%%%%%%%%%%%%
\subsection{EMC}
\noindent The battery pack, as an individual component with connectors and interfaces, including all cells and electronics, has acceptable EMC behavior, as defined in relevant standards. The different types of emissions are limited to a certain level for different frequency ranges. Since the battery pack has a metallic enclosure, it is usually critical to have good connectors for EMC behavior. There may be some shielding and the addition of an XY capacitor to enhance this behavior. EMC filters are commonly used in different parts of the system. For example, it is common to have an EMC filter in the DC output of a pack. However, the most significant source of electromagnetic interference (EMI) in EV, particularly concerning homologation challenges, is the on-board charger. Due to its high switching frequencies and power levels, the on-board charger generates both conducted and radiated emissions that can interfere with other vehicle systems. As a result, special attention must be given to its EMC design, including proper shielding, filtering, and compliance validation. To ensure adherence to UN ECE Regulation No. 10, which governs the electromagnetic compatibility of road vehicles, the emissions and immunity performance of the on-board charger and other key vehicle components must be tested against standardized limits. This regulation specifies acceptable EMI levels and testing procedures to ensure the safe and interference-free operation of electrical systems.

Practical EMC testing conditions follow standardized methodologies, such as those outlined in ISO 11451, serves as a foundation for program owners, regulators, and accreditation bodies to evaluate and acknowledge the competence of validation and verification organizations, including ISO 11451-1 \mbox{\cite{ISO11451_1}}, ISO 11451-2 \mbox{\cite{ISO11451_2}, ISO 11451-3 \cite{ISO11451_3}}, ISO 11451-4 \mbox{\cite{ISO11451_4}}, ISO 11452, which is a collection of international standards and guidelines for testing the immunity of automotive electrical components to narrowband radiated electromagnetic interference from external sources, including ISO 11452-1 \cite{ISO11452_1}, ISO 11452-2 \cite{ISO11452_2}, ISO 11452-3 \cite{ISO11452_3}, ISO 11452-4 \cite{ISO11452_4}, ISO 11452-5 \cite{ISO11452_5}, ISO 11452-7 \cite{ISO11452_7}, ISO 11452-8 \cite{ISO11452_8}, ISO 11452-9 \mbox{\cite{ISO11452_9}}, ISO 11452-10 \mbox{\cite{ISO11452_10}}, and ISO 11452-11 \cite{ISO11452_11}, ISO 7637, which is relates to 12- and 24-volt electrical systems, consisting of ISO 7637-1 \mbox{\cite{ISO7637_1}}, ISO 7637-2 \mbox{\cite{ISO7637_2}}, ISO 7637-3 \mbox{\cite{ISO7637_3}, and ISO 10605 \cite{ISO10605}}, and CISPR (CISPR12 \mbox{\cite{CISPR12}}, CISPR25 \mbox{\cite{CISPR25}}, and CISPR36 \mbox{\cite{CISPR36}}), which define measurement techniques for both conducted and radiated emissions, and related regulating electromagnetic interference in electrical and electronic devices as part of the International Electrotechnical Commission. These tests utilize shielded enclosures, bulk current injection, and reverberation chambers to assess compliance and more effectively pinpoint sources of EMI.

Pack radiation can be measured practically using the recommended setup, as stated in the standard. However, the main challenge is correlating EMC behavior with individual components. For example, if there is a peak in radiation at a certain frequency, it is challenging to find the exact source of interference due to the complex nature of EMC at the pack level, practical measurement and trial and error are widely used to enhance EMC performance. There are many standards that the vehicle should fulfill, which depend on each country. The standards describe the intended behavior in different frequency ranges. However, automakers may have their internal requirements that are usually tougher than other standard requirements. 
Usually, pack-level EMC is handled by the automaker, but at the component level, the supplier will have the main responsibility. For example, the BMS supplier should deliver the part with predetermined EMC levels or standard requirements. The automaker can then assess the EMC at the pack level and decide whether additional mitigation strategies are needed.
Occasionally, practical trial and error is insufficient at the pack level, so more research is required on this subject. In addition, the vehicle is examined for EMC testing, and then, the results are correlated to the subsystem level. The spectrum analysis is important since the detection of the violation's source is concerned. Afterward, a mitigation strategy is prepared and implemented.
%%%%%%%%%%%%%%%%%%%%%%%%%%%%%%%%%%%%%%%%
\subsection{Measurement and Diagnostics}
\noindent The battery pack shall report its state of charge and the status of the system components to the vehicle controller. In addition, in some cases, such as an overcurrent, the pack should be able to act appropriately. 
A combination of cells constitutes a module and a combination of modules forms a pack. The module can have a structure similar to 4S3P, which means that three cells are connected in parallel, and four of these parallel connected cells are connected in series to form a 12~V battery module. 
Since 2010, there has been a significant challenge in the  EV industry related to the lack of a common communication standard for battery packs and BMS across different manufacturers. The absence of a universal standard has led to difficulties when connecting various EV battery systems to diagnostic tools and other vehicle systems. Also, it effectively limited the progress of diagnostic tools that can connect with battery packs from various manufacturers. Nevertheless, substantial progress has been made, and tools like AUTEL play a key role in diagnosing a wide range of EVs \mbox{\cite{AUTEL}}. AUTEL distinguishes itself by its capability to connect with a wide variety of battery packs, because of its wide range of adapters which are designed to support different connectors and plugs. Although improvements in communication standards are ongoing, the industry still encounters significant challenges in achieving universal interoperability across the entire EV battery system. For each battery module, all cell voltages and temperatures typically need to be measured in the BMS. 
Typically, there is a current sensor for the entire pack that measures the final output current. This means that there is no need to measure all modules separately, or even measure the individual cell currents that are connected in parallel. The entire pack voltage and current should be measured with an accuracy of approximately 1-3\% which provides the possibility of calculating the pack power with an accuracy of approximately 3-5\%. Having different accuracy values at different operating points is common. For example, for measurement and estimation, the required accuracy should be higher than the accuracy needed for protection. 
If the battery pack and consequently the energy, is calculated more accurately, it is equivalent to a greater driving range with one charge. Usually, a 5-10\% capacity reserve in a battery is considered for EVs. The new trend is to reduce this number, and companies are now trying to reduce this reserve capacity by better estimating the system. However, extensive testing, data collection over the years, and aging and environmental conditions, including driving behaviors, provide a more accurate estimation of battery energy over its lifetime. Therefore, it is possible to reduce the reserve energy of the pack, which is an important selling point for customers.
In addition to the required measurement accuracy, the system should be able to report its operating status, including faults. If a fault occurs, it is recorded in the system. The diagnostics involve knowing the system status, including subsystems, as much as possible. For example, if the main contactor is welded, the system can detect the fault. In addition, there is a large action list to identify what action should be taken in the case of a failure. For diagnostics, one can mention all cell voltages, main contactor and pre-charge contactor status, auxiliary supply status, current sensor status, connector status (if they are connected or disconnected), and some other similar signals. 
The diagnostics are nearly correlated with the automotive safety integrity level (ASIL) of the subsystem. As a result of the required ASIL level, there could be different measurement solutions from a mix of sensor and software estimation to a level where a redundant sensor is used. This is discussed in the Safety section. For each measurement point, the required ASIL level was part of the requirement. The ASIL level was determined after performing a full system safety analysis.\par
%
%%%%%%%%%%%%%%%%%%%
\section{Battery Pack Interfaces}
\begin{figure}
\centering
\includegraphics[width=1.5in]{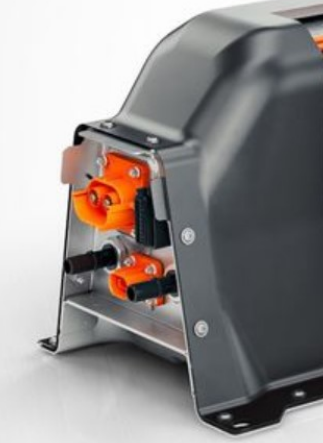}
\caption{Battery pack electrical and coolant connectors.}
\label{Int}
\centering
\vspace{0.4cm}
%\end{figure}
%\begin{figure}
\includegraphics[width=1.5in]{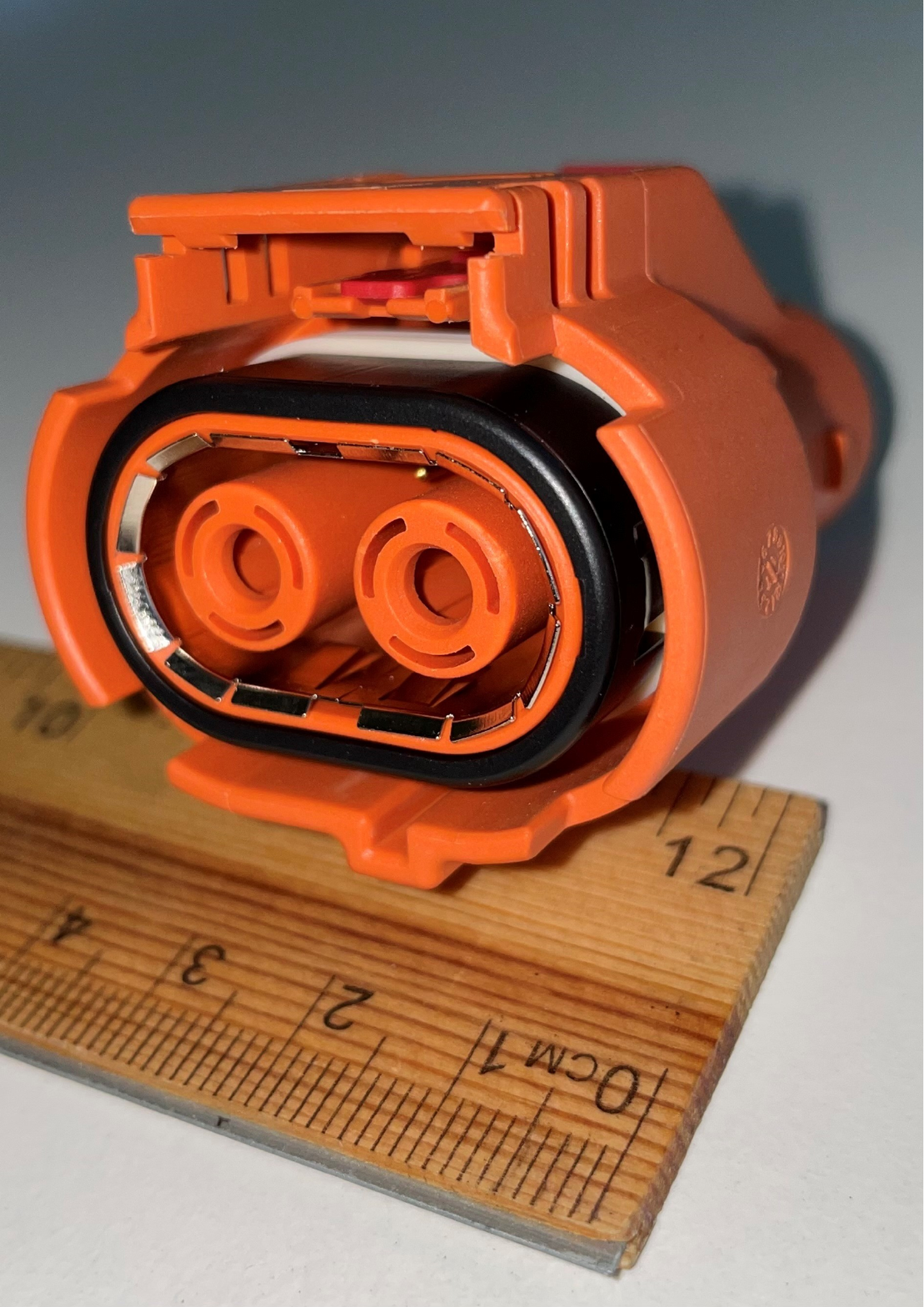}
\caption{Automotive electrical power connector from Amphenol.}
\label{Con}
\end{figure}

\noindent The battery pack, as a subsystem of the vehicle, is mechanically and electrically connected to the vehicle and can communicate with other subsystems. Fig.~\ref{Int} shows an example of the connectors of the battery pack.\par 
%%%%%%%%%%%%%%%%%%%%%%%%%%%%%%%%%%%%%%%%%%
%%%%%%%%%%%%%%%%%%%%%%%%
%%%%%%%%%%%%%%%%%%%%%%%%%%%%%%%%%%%%%%%%%%
\subsection{Mechanical Interfaces}
\noindent The battery pack is an independent subsystem at the vehicle level that was tested separately. The unit should be able to operate under different electrical and environmental conditions considering safety concerns and regulations. Hence, the battery unit is enclosed by a metallic enclosure that can survive different test cases such as vibrations. 
A mixture of water and glycol is commonly used for cooling. The cooling circuit including the pumps and heat exchangers, was outside the pack. However, there is an inlet and outlet connection for cooling. There may be temperature or pressure sensors for the coolant, which are usually outside the battery pack. There are different automotive class connectors for the battery cooling system, which can be referred to as products from Araymond Company. The coolant flow is controlled by a pump, and a capacity in the range of 2-10 liters per minute (LPM) can be assumed for the battery pack. The coolant flow rate in a battery pack is usually less than the coolant flow rate in the motor and inverter. 
The latest design of battery packs is converging towards a flat pack design located under passenger seats. The unit is connected to the vehicle chassis, and the mechanical installation is an important part of the vehicle structure. There is an acceptable level of mechanical guidance for installing a heavy pack in the right spot. 
It is worth mentioning that access to a pack might be required for maintenance or repair purposes. At present, if a single cell fails and one needs to change it, it is not an easy process. The entire pack is detached from the vehicle, and it can be challenging to replace a cell. This is a new aspect of pack design that is under investigation by automakers, especially for aftermarket purposes. 
Access to the connector should be such that the battery pack provides adequate support or mechanical protection. For example, if a vehicle is under rain or driving in a waterway, the connector orientation and mechanical setup should be such that the water, can not be stored close to the connectors. In addition, the connector should have some degree of mechanical protection to survive in extreme cases or during a crash. As shown in Fig.~\ref{Int} the connectors are well supported by the provided mechanical extension in the pack enclosure. To access a manual service disconnect (MSD), it is necessary to use a tool to access the switch. Hence, all requirements and regulations should be considered in pack design.
\subsection{Electrical Interfaces}
\noindent There are three types of electrical interfaces for a battery pack: power, signal, and MSD. The battery pack may have one or more main outputs to be connected to the loads and charger. In the new designs, the battery pack output is connected to a junction box or distribution box to supply loads with fuse protection. Hence, appropriate power connectors with disconnect diagnostics must be used.
Each power connector has an auxiliary connection with a few amplifier rating values that are used to detect connector disconnection. If a cable is disconnected, its auxiliary circuit loop is broken, and the system can detect the disconnection. In an automotive power connector, a small connection can be seen for this purpose, as can be seen in Fig.~\ref{Con}. 
In addition to power connectors, there is a signal connector that includes communication signals, such as CAN or LIN. In addition, a crash signal informs the battery unit about the occurring crash. In this case, the battery pack takes safety action and disconnects the internal contactors. This connector is an appropriate automotive-class connector. 
As mentioned earlier, an MSD is used to break the battery pack circuit for maintenance. This is used when technicians intend to open a pack. To reach the MSD as a manual power switch, a small metallic enclosure is usually opened by a tool. 
%%%%%
\section{Environmental Requirements}
\noindent It is desirable to have a maintenance-free operation and a longer service time. Safety is an important feature that should be tested under different operational conditions, as discussed in the next section. Hence, to achieve these goals, well-regulated standards have been designed and are in place as environmental standards. The idea was to test the components under different environmental conditions to verify the effectiveness of the product. Mechanical shock and vibration, heat and humidity, chemical contamination, and other tests were conducted. The tests are performed at the vehicle and subsystem levels. Usually, automakers are responsible for vehicle-level testing and suppliers are responsible for component-level testing.
At the vehicle level, a widely accepted regulation has been developed under the United Nations (UN), which is called ECE 100. A vehicle with an electric powertrain should comply with this regulation. In addition, the battery pack was covered by this standard\cite{ECE100R2}. This standard also covers safety concerns. At the component level, battery packs or parts such as the BMS and ISO 16750 series (road vehicles, -- environmental conditions, and testing for electrical and electronic equipment) are very common \cite{ISO16750-1,ISO16750-2,ISO16750-3,ISO16750-4,ISO16750-5}. ISO 16750 was originally developed for engine-based vehicles but is still used for hybrid and  EVs. More widely accepted EV-dedicated standards are expected shortly. 
The second part of ISO 16750 considers the electrical specifications and is called the electrical load in the standard \cite{ISO16750-2}. For example, a BMS board can be the subject of this test. The power supply variation, reverse polarity connection of the device, open-circuit tests, short-circuit protection, and insulation resistance are part of this section. 
The third part of the standard is devoted to the mechanical aspects \cite{ISO16750-3}. Vibration, mechanical shock, and stress are the subjects of this standard. 
Part four of the standard is called climatic loads, which concern temperature, humidity, dust, corrosion, salt water, ice tests, and similar topics. The most difficult part of the test is the combined humidity and vibration test, which is very challenging. Depending on the installation location inside the vehicle, different maximum values were tested. For example, the climate temperature range is $-45^0C$ to $50^oC$ but if a component is close to the internal combustion engine, the maximum temperature will be increased to $80^oC$ for testing. Another case is when the vehicle is under direct sunlight for a while, and the temperature is increased more than in the environment. Hence, the standard for the value of temperature and cycle testing considers the class of the component that should be referred to. 
The final part of the ISO 16750 standard is dedicated to chemical issues \cite{ISO16750-5}. Electronic boards or devices are subject to different chemical materials for a predetermined time or cycle. The concept is that the component should perform as stated in the standard during or after the test. In most cases, a visual check is required to detect possible damage to a component.
As mentioned in this section, there are different types of environmental tests to ensure the performance of the component for different cases. For example, the pack drop test, water immersion, and similar tests are part of the ISO 1670 series or comparable standards. 
In addition, other standards regulate other aspects, such as production or labeling requirements, or do not use toxic materials. For example, one can refer to lead-free soldering materials or bans on cadmium materials in battery cells.
%%%%%%%%%%%%%%%%%%%%%%%%%%%%%%%%%%%%%%%%%%%%%%%%%%%%%%%%%%%%%%%%%5 
%%%%%%%%%%%%%%%%%%%%%%%%%%%%%%%%5
\section{Safety}

\noindent Safety is an important feature of EVs, as an emerging technology, which play a pivotal role in the EV's performance. Thermal runaway is one of the major concerns of safety, which typically occurs due to excessive heat, internal short circuits, overcharging, and mechanical damage to the battery cells. If these things happen, it causes the failure of the battery pack, fire, or even explosion, which poses a significant hazard to the EV and passengers. In this regard, to mitigate the possible problems associated with thermal runaway, the BMS is used, which is designed to maintain safe operating temperatures of the battery pack,  preventing overheating and minimizing the probability of thermal runaway events. Proper temperature regulation is a crucial factor in the performance, longevity, and safety of battery packs.
There are various technologies used within BMS to control the temperature of the battery pack \cite{R3_Comprehesive}:
\begin{itemize}
    \item Active and Passive Cooling: Active cooling systems use pumps and liquid coolants to remove excess heat from the battery cells, ensuring that cells operate in the safe temperature range. On the contrary, passive cooling systems use natural convection, heat pipes, or other methods to dissipate heat flow without the use of external power sources. Nevertheless, both active and passive cooling mechanisms are essential in preventing overheating of the battery pack. These systems prevent the starting thermal runaway and ensure that the battery temperature is in the safe thermal limits during high-demand scenarios, like fast charging or discharging with high power   \cite{R3_thermal,R3_Comprehesive,R3_development}.
    \item Phase-Change Materials (PCMs): PCMs are substances which absorb heat while they change phase from solid to liquid at a specific temperature. Some BMS use the PCM to help absorb excess heat during high charge or discharge C-rates. By doing so, a more uniform temperature across the battery pack can be obtained, which prevents the hot spot due to overheating. When the temperature rises, the PCM absorbs the energy and transitions to a liquid state, thereby stabilizing the battery pack's temperature. When the temperature falls, the material solidifies, releasing stored heat. This capability can extremely enhance the safety and efficiency of the EV battery packs \cite{R3_development}.
    \item Immersion Cooling: Immersion cooling involves submerging the battery pack inside a thermally conductive liquid that directly absorbs heat from the cells. As compared to air cooling systems, this method provides more uniform and efficient cooling. Immersion cooling is especially useful for high-power applications, in which fast temperature fluctuations are more likely. By submerging the cells in a liquid coolant, the heat is more effectively transferred from the cells to the coolant, reducing the possibility of thermal runaway or having the hot spot \cite{R3_thermal}.
\end{itemize}
Besides these technologies, BMS also plays a crucial role in safety by monitoring the battery's voltage, doing corrective action when abnormal conditions are detected, and other critical parameters. For example, the BMS can enable the cooling system or limit charging rates to prevent overheating if it is required. There have been several reports of EV accidents~\mbox{\cite{Crash}}. There are different safety layers and standard regulations from the vehicle level to the sub-component level. In response to fulfilling standard regulations and achieving a safe system, the ISO 25265 series has been established, followed by automakers and suppliers
\cite{ISO26262-1,ISO26262-2,ISO26262-3,ISO26262-4,ISO26262-5,ISO26262-6,ISO26262-7,ISO26262-8,ISO26262-9,ISO26262-10,ISO26262-11,ISO26262-12}. The topic is briefly reviewed in this section, and it is expected that standards or relevant references will be visited for more information.\par 
\subsection{UN ECE Regulation No. 100}
\noindent UN ECE Regulation No. 100 \cite{ECE100R2} is an international standard for harmonizing vehicles with electric powertrains and energy storage systems. The safety of vehicles and energy storage are addressed in this regulation at the vehicle level. 
The first part of the standards concerns the vehicle's electrical safety requirements. Thus, protection against electrical shock should be secured. There are some mechanisms to achieve safety to avoid direct contact with high-voltage parts by insulating or packaging, marking high-voltage areas, securing connectors and connections to avoid shock even during service operation, and providing sufficient insulation resistance between the high-voltage part and chassis. "For example, a device connected to a high-voltage bus has an insulation resistance greater than 1 megohm. However, some details may alter this limit.
The second part of this standard concerns the safety of the energy storage system, which can be a battery or a fuel cell unit.
Energy storage systems include vibration, thermal shock and cycling, mechanical shock, mechanical integrity, fire resistance, external short circuit protection, overcharge protection, over-discharge protection, and over-temperature protection. The acceptance criterion is that the battery will not exhibit any fire, explosion, electrolyte leakage, or insulation resistance reduction. However, one needs to see the detailed test procedures as explained in the standard. However, brief descriptions of some tests are provided below.\par 
%%%%%%%%%%
%%%%%%%%%%
%%%%%%%%%%
\begin{itemize}
  \item Vibration: The battery pack was subjected to vibration with a sinusoidal waveform with a frequency sweep between 7~Hz and 50~Hz in 15 min. The test was performed for three hours.
  \item Thermal shock and cycling: The battery back environmental temperature was changed from $-40^o$C to $60^o$C for 24~h. Each time, the temperature was established for six hours. The idea is to emulate environmental conditions.
  \item Mechanical shock: The purpose of this test is to verify the battery pack under an inertial load during a crash. The pack is subject to different longitudinal and transverse accelerations, as indicated by the standard.
  \item Mechanical integrity: This test evaluates the mechanical forces acting on a pack subject during a crash. A force around 100~kN was applied to the pack at 100~ms. 
  \item Fire resistance: The battery pack should be resistant to external fires. Hence, the pack will be subject to the fire and will survive, or the driver will have enough time to evacuate!
  \item External short-circuit protection: The battery pack is subjected to a short circuit. The pack should operate after a short circuit.
  \item Overcharge protection: The pack will be overcharged, and the protection will stop charging. Overcharging is a critical condition that can result in thermal runaway.
  \item Over-discharge protection: The over-discharge is tested to evaluate the protection mechanism.
  \item Over temperature protection: The purpose of this test is to evaluate the internal overheating of the pack under different conditions, such as the loss of the cooling system.
\end{itemize}
%%%%%%%%%%%%%%%%%%%%%%
\subsection{ISO 6469: Electrically propelled road vehicles — Safety specifications}
\noindent Another important safety standard is ISO 6469 \cite{ISO6469-1,ISO6469-2,ISO6469-3,ISO6469-4}, which has four parts. This standard addresses the safety of energy storage systems and electrically driven vehicles. Part four of the standards discusses post-crash situations. There are some similarities and differences between this standard and US ECE 100.R2.
The safety of human beings and livestock considering the electrical shock is mentioned in the IEC 60479 standard \cite{IEC60479_1,IEC60479_2}. The stored energy in a point of a circuit shall be less than 0.2~J to be considered safe. Hence, there are some tests and requirements to ensure that the stored energy at any point in the high-voltage circuit does not exceed this value. This sets a limit on the total available Y capacitance, which is the capacitance between a circuit point and the safety earth, that is, the chassis in the vehicle case. The X capacitor is between the positive and negative ports of the battery, and the Y capacitance is between the positive and negative terminals of the chassis. The X and Y capacitors improve the EMC performance. However, their upper values are limited by safety. The maximum Y capacitance of a point can be calculated based on the system voltage and limit of the stored energy as follows: 
\begin{equation}
\label{J}
W=\frac{1}{2}CV^2
\end{equation}
where $W$ is the stored energy, $C$ is the capacitance and $V$ is the voltage, respectively. 
Another important aspect of the standards is the system behavior after a crash or even disconnection of a connector. The voltage level was reduced to a safe level under these conditions. After a crash, people may come to the rescue, and there should not be a risk of electrical shock. For this reason, a time of approximately 1-2 minutes is considered in the standard to reduce the voltage to a safe value. In the case of connector disconnection, this time is approximately a second. Hence, especially in new designs, the aim is to achieve a safe voltage level in seconds. For example, if the motor winding is short-circuited by the inverter action, a DC bus capacitor will be deployed, and the voltage will be reduced quickly. This is called an active discharge. Another way to reduce the voltage is to add a large resistance value to discharge the DC bus capacitor in a couple of minutes, which is called passive discharge.
%%%%%%%%%%%%%%%%%%%%%%%%%%%%%%%%%%%%%%%%
\subsection{ISO 26262: Functional Safety Standard}
\noindent ISO 26262 (Road vehicles — Functional safety) \cite{ISO26262-1,ISO26262-2,ISO26262-3,ISO26262-4,ISO26262-5,ISO26262-6,ISO26262-7,ISO26262-8,ISO26262-9,ISO26262-10,ISO26262-11,ISO26262-12} is an adaptation of IEC 61508 \cite{IEC61508} that provides a framework for identifying the potential risks of software and hardware failures in automotive electronics. The standard has 12 parts that cover the entire life cycle of electronic components, from the early concept phases to aftermarket activities.
An important concept for quantizing risks is the automotive safety integrity level (ASIL). Risk is divided into four levels: A, B, C, and D. However if a signal is not ASIL, then it is QM, quality management. A hazard assessment and analysis are performed, and based on the severity, exposure, and irreconcilability of the signal, an appropriate ASIL level is assigned to a signal. For a QM signal, a simple quality check is sufficient, whereas an ASIL D signal requires special care that can be a hardware/software-level redundancy. It is worth mentioning that there is no technical solution to a particular problem in ISO 26262, but it is a systematic way to manage complex automotive electronics.
In the first step, hazard analysis and risk assessment (HARA) were performed. Based on the results, safety goals were selected. For example, for a battery pack, the HARA results might be due to the risk of fire on overcharging or after a crash. Subsequently, the safety goal (SG) can be determined, identified, or selected. For example, for an automotive battery pack, part of the safety goals can be as follows\cite{Marcos,Li}:
\begin{itemize}
  \item SG1: battery shall not get fire, ASIL D
  \item SG2: battery shall not emit toxic gases, ASIL D
  \item SG3: High voltage shock shall be avoided, ASIL D
   \item SG4: battery shall not allow over-charge, ASIL C
   \item SG5: battery shall not allow over-discharge, ASIL B.
    \end{itemize} 
As can be seen from the list, the ASIL requirement was also identified. In response to the safety goals, another document will be created as a safety requirement. The safety requirements document contains more details. The safety requirements are sent to the supplier as inputs to the design phase. The developed hardware satisfies safety requirements as well as other requirements. The supplier reports how the safety requirements are fulfilled by performing different types of analyses and tests. For example, it is common practice to perform a failure tree analysis to determine the consequences of a single failure. Different types of failure mode and effect analysis (FMEA) will be performed at different stages of development. For example, DFMEA was performed in the design stage. Another document is a worst-case analysis that shows how the system or part of the system behaves. For example, if the resistance increases by 5\%, what happens in the circuit? As expected, these reports can be long in many cases. A BMS circuit can have an FMEA with more than 4000 fault cases. 
All parties were involved in safety work. The automaker is responsible for the system, while the suppliers are working at subsystem levels. Component manufacturers are also involved in this process. For example, if there are two ASIL D signals, it is not possible to connect them to any input of the microcontroller. There is a safety document for the microcontroller, and the pin assignment shall be according to the microcontroller manufacturer's recommendation. If two ASIL D signals use an internal multiplex inside the microcontroller and if the multiplex faces a fault, then ASIL D may not be feasible. Consequently, to reach the required ASIL D in a part of a system, a full analysis of the entire system is required, which can be very challenging. There are additional steps and documentation in the process that are not mentioned here to limit the scope.
%%%%%%%%%%%%%%%%%%%%%%%%%%%%%%%%%%%%%%%%%%%%%%%%%%%%%%%%%%
\section{Battery System Architecture}
\noindent So far, different requirements for automotive battery packs have been described. This section presents the automotive battery pack structure, components, and performance. The idea is to reflect the latest developments and near-future trends by considering the available components in the market. 
A battery capacity of approximately 100~kWh is the typical capacity for many EVs considered here. For a battery system with a 400~V nominal voltage, the battery pack has approximately 250~Ah. The pack is a series and parallel connection of cells. Hence, it is important to provide information regarding the available cells.
%%%%%%%%%%%%%%%%%%%%%%%%%%%%%%%%%%%
\subsection{Cell Technology}
\begin{figure}
    \centering
    \includegraphics[width=0.8\linewidth]{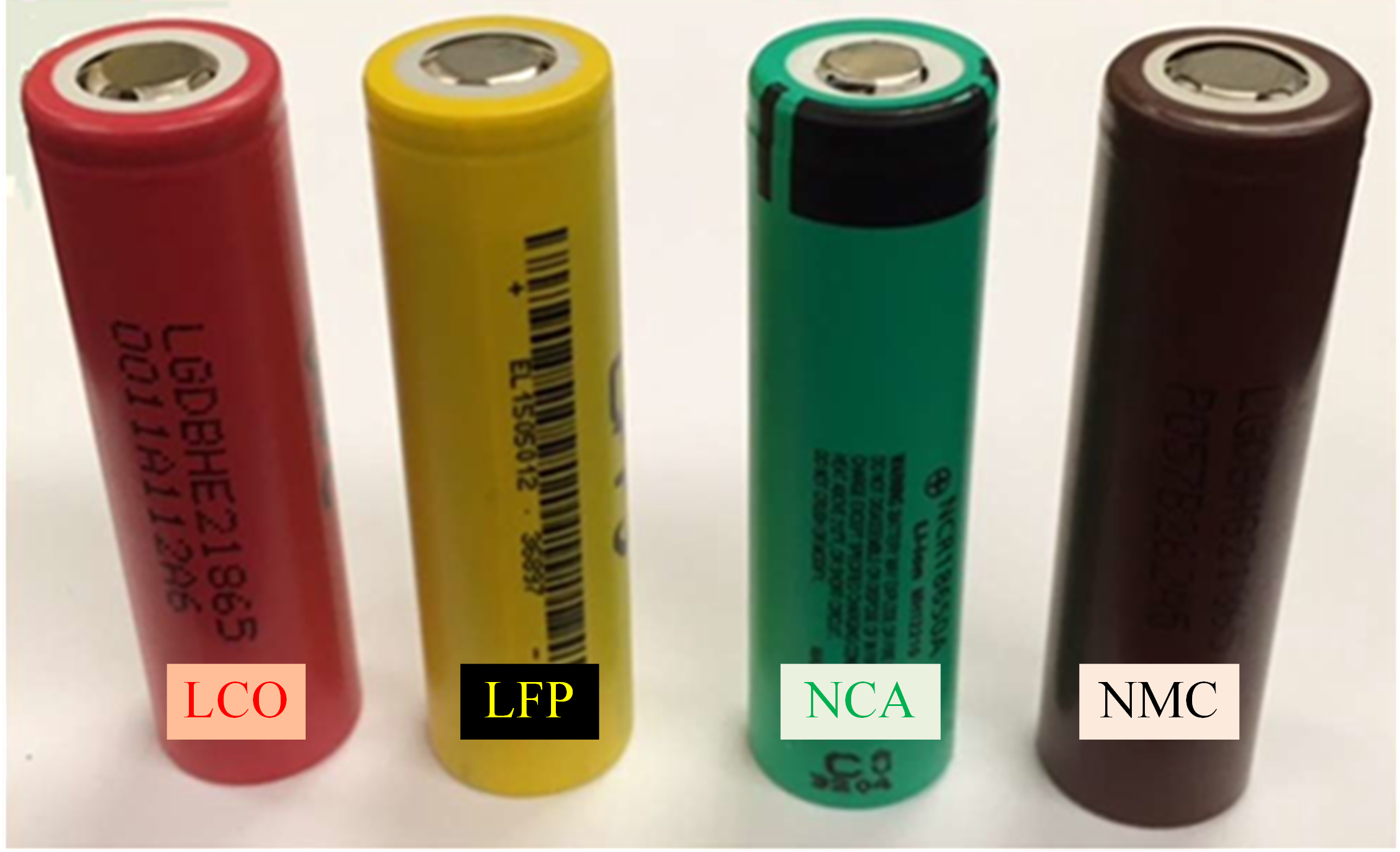}
    \caption{Battery cells based on Cathode types. (a) LCO. (b) LFP. (c) NCA. (d) NMC \cite{Barkholtz}.}
    \label{fig:LFP}
\end{figure}

\begin{figure}
    \centering
    \includegraphics[width=1\linewidth]{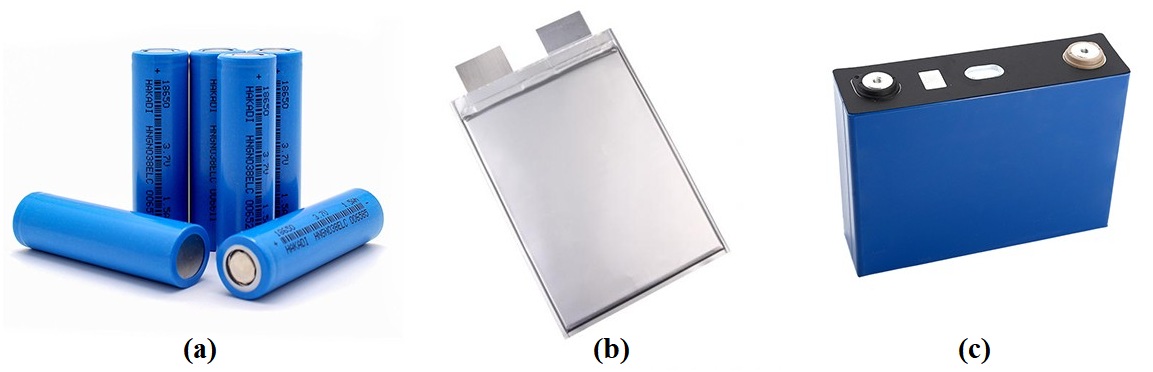}
    \caption{Battery cell packages. (a) cylindrical. (b) pouch. (c) prismatic.}
    \label{fig:batteries}
\end{figure}

\noindent Li-ions are the dominant chemical species in the automotive battery cells. There are different types of common li-ions batteries, such as Lithium Cobalt Oxide (LCO), Lithium Iron Phosphate (LFP), Lithium Nickel Cobalt Aluminum Oxide (NCA), and Lithium Nickel Manganese Cobalt Oxide (NMC), to name but a few. Fig. \mbox{\ref{fig:LFP}} shows the LCO, LFP, NCA, and NMC battery types, which are different in the cathode types and have different nominal voltage, mass, energy density, and so on. The characteristics of these batteries are presented in Table \ref{tab:batteries}. 
\begin{table}[!b]
    \centering
    \caption{Li-ion battery technologies characteristics \cite{Barkholtz,Larijani_book}.}
    \begin{tabular}{c|c|c|c|c}
       Battery & LCO & LFP & NCA & NMC \\
       \hline  
       Cathode & $\mbox{LiCoO}_2$ & $\mbox{LiFePO}_4$ & $\mbox{LiNi}_{0.8}\mbox{Co}_{0.15}\mbox{Al}_{0.05}\mbox{O}_2$ & $\mbox{LiNi}_{0.8}\mbox{Mn}_{0.15}\mbox{Co}_{0.05}\mbox{O}_2$ 
       \\
       Nominal capacity ($Ah$) &2.5&	1.1&	2.9&	3
       \\
       Nominal voltage ($V$)&	2.5&	1.1&	3.6&	3.6
       \\
       Maximum discharge &	20&	30&	6 & 20
       \\
       current ($A$) & & & &
       \\
       Mass ($g$)&44.9&	39.4&	44.8&	44.9
       \\
       Volumetric energy density &533.4&	212.1&	676&	612.1
       \\
       ($Wh/L$) & & & & 
       \\
       Specific energy density&195.8&88.9&243&224.8
       \\
       ($Wh/kg$)& & & &
       \\
       Operating temperature &0 to 50&-20 to 55&0 to 45&0 to 50
       \\
       range under charge ($^{\circ}C$)& & & &
       \\
       Operating temperature &-20 to 55&-30 to 55&-20 to 60&-20 to 55
       \\
       range for discharge ($^{\circ}C$) & & & &
    \end{tabular}
    \label{tab:batteries}
\end{table}
NMC and LFP cells are widely used in automotive applications. The LFP has a lower voltage but better temperature stability \cite{Americans}.  
Battery cells are available mainly in three packages: pouch, cylindrical, and prismatic, as shown in Fig. \ref{fig:batteries}. The pouch is easier to produce and to cool in mass and is used by most automakers. In contrast, Tesla uses cylinder cells that require a more complicated cooling structure and better safety. 
As mentioned earlier, cell chemistries are very different, and it is difficult to classify them accurately. Examples of main cell manufacturers for automotive applications are LG, CATL, and Panasonic. More companies and new companies, such as Northvolt, will emerge as well.
The ideal cell should be sufficiently high to provide the required ampere. For the above-mentioned 100~kWh pack, the ideal single-cell capacity is 250~Ah. However, the available cell in the automotive industry is approximately 80~Ah for the pouch type. Hence, three cells were be connected in parallel to provide the required $Ah$. Pouch cells with a capacity of approximately 200~Ah are under development, and it is expected to see them in a couple of years.  
%
%%%%%%%%%%%%%%%%%%%%%%%%%%%%%%%%%%%%%%%%%%%%%%%
\subsection{Battery Module}
\noindent The battery pack includes several battery modules that can have a voltage of $12~V$ or $24~V$. Higher voltage levels are expected to be seen as well. Usually, modules are connected in series to provide the required voltage and capacity. If we consider the pouch cell in the above example, some cells can be connected in parallel; for example, 3. Then, four of these units are connected in series to reach the 12~V or eight will be in series for 24~V units. 
Because the module voltage level was low, it was easier to produce and test. The cell voltages and some temperature points were monitored at the module level, as explained in the BMS section. The module is the building block of a battery pack. Normally, there is only one type of battery module; however, the number of modules can vary in a pack. 
As an example, consider a pouch cell with dimensions of $10 \times 55 \times8$ cm, a capacity of 80~Ah that has a module with a $4S3P$ configuration. This is a 12~V battery module with 240~Ah capacity. Liquid cooling with a cold plate is assumed for this design. The capacities of the cylindrical cells were much lower than those of the pouch cells. Hence, there were more cells. For example, if 18650 cells with a capacity of 3.5~Ah are used, the $4S60P$ configuration should be used in a similar module. For the cooling of a cylindrical cell, water tubes can be used when the cells are in contact with the tube. One can refer to \cite{Full}, which provides a technical comparison between the cylindrical, pouch, and prismatic cells for automotive battery packs.
%%%%%%%%%%%%%%%%%%%%%%%%%%%%%%%%%%%
\subsection{Battery Electric Power Architecture}
\begin{figure}
\centering
\includegraphics[width=2.0in]{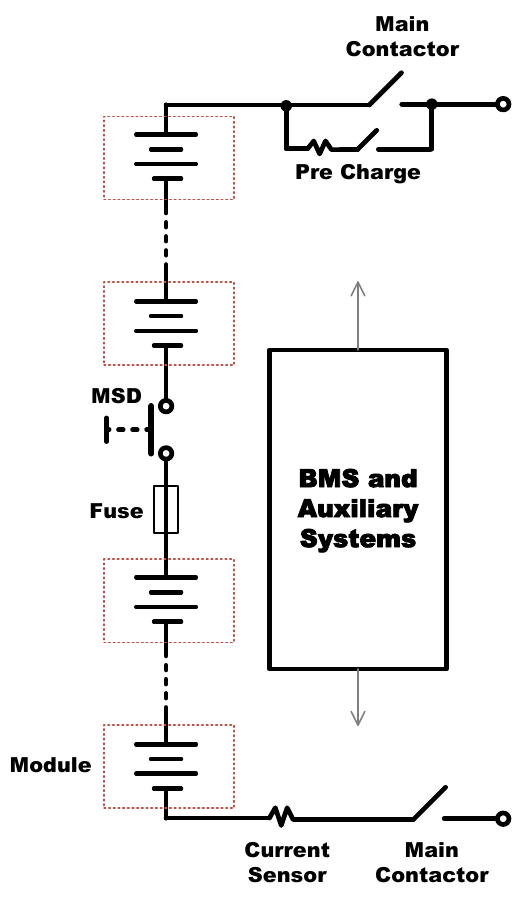}
\caption{Battery pack electrical circuit schematic.}
\label{Arc}
\end{figure}

\noindent The power diagram of an automotive battery pack is shown in Fig.~\ref{Arc}. This structure is widely used in many vehicles and is the standard configuration. Despite its simple structure, the system is complex, considering all safety requirements. For example, there are two main contactors for redundancy. The system should be able to identify whether the contactor is welded as an example. This section discusses these aspects. 
In the middle part of the pack, there was a fuse and an MSD. After finishing the pack, if a technician needs to perform maintenance or any other activities, the MSD should be removed first. In this way, the battery pack is divided into two parts: if the MSD is in the middle position, the high voltage of each part is reduced by half, making it safer to work with. The fuse should be sized so that it can operate in extreme cases, such as short circuits, where the contactors are not fast enough or can not disconnect the load.  
%%%%%%%%%%%%%%%%%%%%%%%%%%%%%%%%%%%%%%%%%%
%%%%%%%%%%%%%%%%%%%%%%%%
A battery may have one or two outputs. In the new designs, the battery output is connected to a junction box where the power is distributed and may be protected by fuses or contactors. For the traction inverter, there may or may not be a fuse in the junction box. For the DC/DC and onboard charger, direct contact is usually used. However, for the fast charger, a contactor and a fuse are used. The onboard charger fuses with the AC side and may fuse with the DC side inside the onboard charger. Usually, there are one or two fuses between each unit and the battery module. Note that the battery pack has a main fuse in place for all the loads and units. Appropriate fuses exist for the compressor and heater. However, the fusing structure might be different in different packs, depending on the safety mechanism and system structure.
The battery pack communicates with the vehicle via a communication bus, which can be a CAN bus, a Linbus, or similar buses. A controller unit performs this task and other tasks, such as measurement, diagnostic, and control. A 12~V supply was connected to this controller as the main power supply. However, there is a redundant power source from the pack itself as well.
A current sensor is used to measure the pack current. The power and energy of the pack were calculated by measuring the pack voltage. Kalman filter estimators are commonly used for power and energy estimation. As mentioned earlier, it is desirable to have good accuracy for the estimation and trends toward higher accuracy numbers like 1-2\%. 
The main contactor includes a pre-charge circuit that the controller controls. The driver of the contactor includes two switches: one on 12~V side and the other on the ground side. Consequently, these drivers are called high-side drivers and low-side drivers. 
The system includes a range of diagnostics for different parts of the system, some of which are discussed here. The 12~V power supply was monitored and reported. Special power controllers have been developed for this task, which are part of the controller. In addition, the system detects the auxiliary power supply status. The contactor's state of health was detected by the system. Essentially, if a contactor is welded or faulted to an open circuit, the system will detect that. Driver faults were also detected. Typically, the voltages of the two sides of the contactor and the current of the coils are measured to perform the required diagnostics. 
A high-voltage interlock loop (HVIL) is a low-voltage system with a small current signal. All major connectors have auxiliary contacts connected to this loop. If one connector is unplugged, the HVIL is broken and the system detects that. The battery controller is also a part of the loop. 
An ASIL level C or D microcontroller exists inside the controller. There are diagnostic features of this part of the system. For example, the Infineon Aurix 297 microcontroller is widely used in BMS. A part of thermal management can also be performed in the BMS. In this case, the liquid flow may be controlled by the pack depending on the operational conditions. For fast charging, there is an issue with busbar temperature because of the extremely high current. This part of the system may also be monitored by the controller. For example, the charging power can be adjusted based on the busbar temperature.
The cybersecurity requirement of battery packs is a new topic currently in place for automotive applications. The system design should be secure against cyber attacks. The associated risk shall be analyzed, and microcontrollers shall be secured in some aspects, such as having secure communication lines between the battery pack and the vehicle, as well as between the modules. 
%%%%%%%%%%%%%%%%%%%%%%%%%%%%%%%%%%%%%%%%%%%%%%%%%%%%%%%%%%%%%%%%%%%%%%%%%%%
\section{Battery Management System (BMS)}
\noindent The BMS in the battery pack has several functions, including monitoring of cells and the temperature at some points, controlling the contactors, monitoring of the pack current, communication towards the vehicle, and additional functions that will be discussed in this section. The idea was to complement the materials provided in the previous sections. BMS plays a critical role in ensuring optimal performance, safety, and longevity of EV battery packs. Among the primary tasks of the BMS are SOC estimation, which determines the remaining charge in the battery; SOH estimation, which evaluates the overall health of the battery; and temperature monitoring, which ensures that the battery operates within safe temperature limits. These parameters are crucial for maintaining efficient and reliable operation.\par 

\subsection{Battery SoC and SoH estimations}

\subsubsection{State-of-Charge (SoC) Estimation}

The SOC estimation determines the remaining charge of the battery cell, which plays a key role to ensure that the battery is neither overcharged nor excessively discharged, both of which can shorten the battery cell life-time. The most common methods for SOC estimation include the following:
\\
\begin{enumerate}
    \item Coulomb-counting Method: This method, can also be called the ampere count method, calculates the battery cell SoC by integrating its current over time \mbox{\cite{Xiong,Larijani}}. Although this method is simple, it can suffer from errors of the accumulation of drift over time, requiring periodic corrections. Furthermore, the initial SoC should be known to accurately compute the SoC over time \cite{Edorado_2}.
    \item Model-Based Methods: This method uses the battery cell model, for which there are recent advancements too, such as those method which are based on Kalman filters, which dynamically adjust to variable conditions. The study by Locorotondo et al. \mbox{\cite{Locorotondo}} presented the use of model-adaptive Kalman filters to estimate SoC more accurately by adjusting the model parameters based on real-time data, like current and voltage measurements. These methods can reduce the error margins which has been seen in simpler techniques, thereby providing more reliable estimates of SoC.
    \item Spectroscopy Methods: These methods use techniques like impedance spectroscopy to estimate SoC by analyzing the internal resistance of the battery and other electrochemical properties \mbox{\cite{spect}}. Although these methods have high precision, actually they are a little complex and require specialized equipment, and are not suitable for real-time implementation.
\end{enumerate}

\subsubsection{State of Health (SoH) Estimation}

SoH evaluates the overall health of a battery which refers to the ability of the battery to perform at the rated capacity and maintain charge retention over time. Model-based approaches, which are similar to those used for SoC estimation, are commonly employed for SoH estimation. These models often incorporate into aging effects and degradation patterns, thereby providing more accurate forecasts of battery performance over their life-time \mbox{\cite{battery_degradation},\cite{Edorado}}.
The temperature is a critical parameter for maintaining battery safety and performance. High temperatures accelerate aging; on the other hand low temperatures can reduce the battery efficiency. The BMS typically includes temperature sensors located at multiple points in the battery pack to monitor temperature fluctuations. Advanced methods for temperature management often deal with combining sensor data with predictive models to optimize the battery's thermal management system in real-time.\\
Despite significant progress, several challenges remain in the accurate estimation of SoC and SoH. For example, temperature variations across cells and the aging effects of battery components can complicate estimation processes. Latest innovations in machine learning and adaptive algorithms show promise in improving the accuracy of these estimations \mbox{\cite{machine_SoC}}. By improving the methods for estimating these parameters, modern BMS contribute significantly to the safety, efficiency, and durability of EV batteries, thereby ensuring that they operate within optimal conditions throughout their lifecycle.
%%%%%%%%%%%%%%%%%%%%%%%%%%%
\begin{figure}
    \centering
    \includegraphics [width=3.5in]{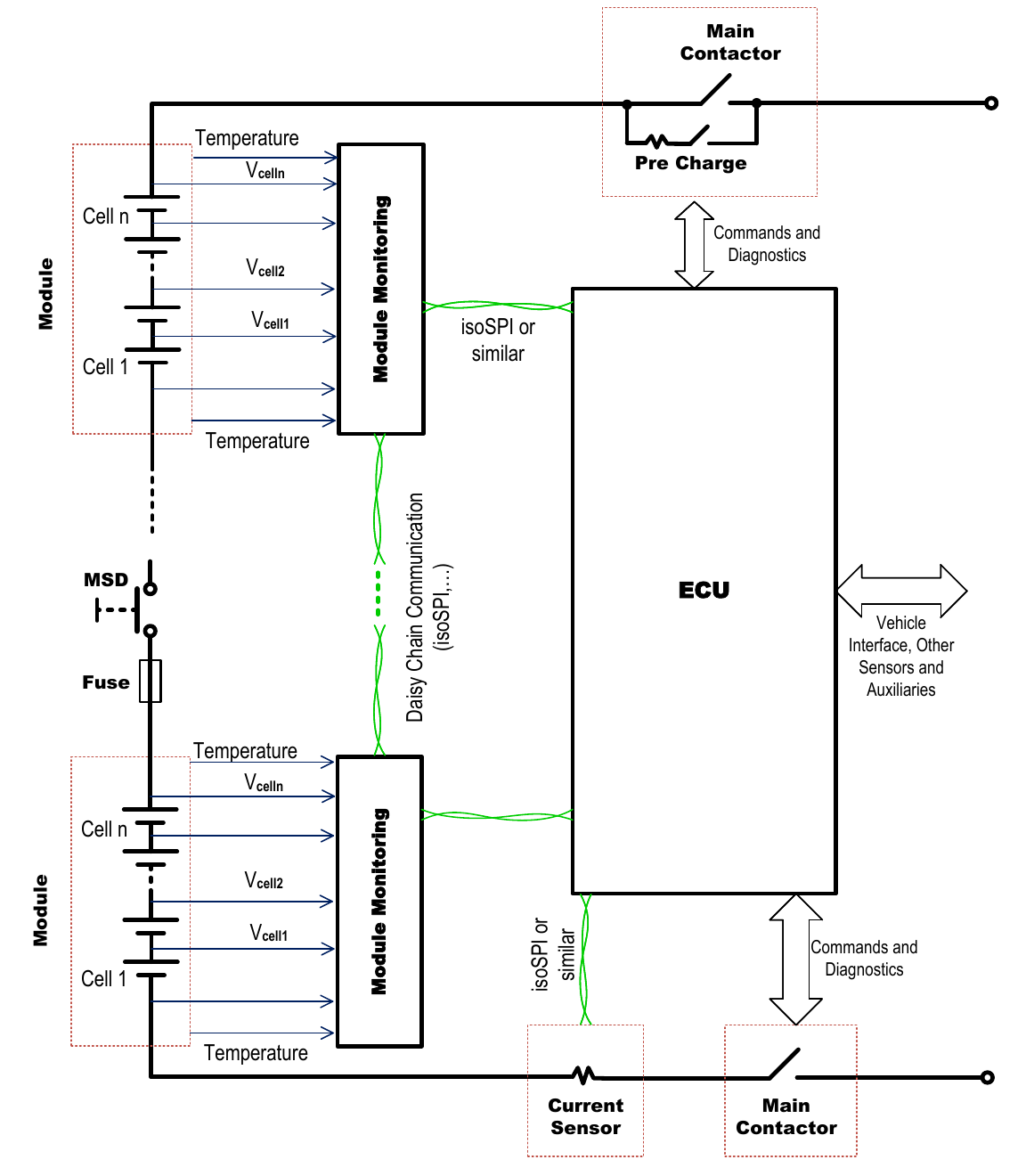}
    \caption{Battery management system (BMS) of a battery pack.}
    \label{ECU}
\end{figure}
%%%%%%%%%%%%%%%%%%%%%%%%%%%%%%%
\subsection{Module Monitoring and Control Unit}
\noindent Fig.~\ref{ECU} shows the BMS, including a daisy communication channel between the modules and the controller. Each module had a monitoring board in which the voltage of the cells and a couple of temperature points of the modules were constantly measured. As mentioned earlier, there are parallel cell connections in the first place. Subsequently, the common voltage was measured. The individual current distribution of cells is not measured owing to the complexity and cost impact. The estimation of the state of charge or available power of parallel-connected cells requires more work, as shown in an example in \cite{Han}.
Temperature measurement was performed inside the module at appropriate points. It can be a couple of points inside the module or more points, including all cells. More points add more complexity, especially when considering the ASIL level of the signals. This adds more limitations to the system. A practical compromise may be the measurement of two points between the cells to ensure that the system does not face a thermal runaway.
Another feature of the module controller unit is cell balancing. There are passive and active cell balancing methods \cite{Naguib,Ghaeminezhad}, in which passive cell balancing is common for passenger cars because of its simplicity and cost. 
Because of the large number of cells in the pack, collecting and processing the required information is a challenging task. The module controller collects all measurements and sends them to the electronic control unit (ECU) via a serial bus, such as isoSPI. In this manner, each module has a pair of wires as the input and another pair of wires as the output. As shown in Fig.~\ref{ECU}, a serial loop was formed, including all module controllers and the ECU. An ECU may have two connection sides. Usually, an ASIL D level is required for communication, and if one cable is disconnected, the second cable can continue the operation. Hence, in this ring configuration, it is possible to reach each node on two paths. 
One important development is the removal of all wiring between the cells and communication and the use of a wireless structure. This activity is ongoing and it is expected to see more products in the market in the near future. One major issue with a battery pack is the integrity of the communication link.
%%%%%%%%%%%
\subsection{Electronic Control Unit (ECU)}
\noindent All pack activities are governed by the ECU. The ECU includes an automotive ASIL D or C auxiliary power supply and communication interface towards module controllers, contactors, sensors, and vehicles. Usually, a double or triple-core microcontroller is used to achieve the required ASIL level.
Moreover, there are two communication channels, 12~V power supply, and HVIL interfaces towards the vehicle. The ECU communicates with the vehicle via the CAN buses, LIN buses, or equivalently. Usually, there are different communication buses for different purposes depending on the vehicle structure.
As mentioned earlier, there is a full range of diagnostics for the different components. The voltage before and after the contactors, the coil current of the contactor, the battery pack current, and some temperature measurement points, such as the PCB temperature, were measured by the ECU. In this manner, the ECU can perform the required diagnostics. As a result, there are different types of actions from a warning message to the driver to open contactors and stop the vehicle.
Another important functionality of ECUs is the monitoring of their insulation resistance. The insulation resistance between the high-voltage positive and negative terminals and the vehicle chassis was measured by the ECU. If the insulation resistance is less than a threshold value, appropriate action is taken. During fast charging, insulation resistance monitoring is bypassed because of the ground connection in the fast charger station. In this case, galvanic isolation was performed in a fast charger station.
%%%%%%%%%%%%%%%%%%%%%%%%%%%%%%%%%%%%%%%%%%%%%%%
\section{Other Topics and New Trends}
\noindent The specific energy and power are important parameters in the cell and pack levels that can be countered as performance indices. As mentioned in the previous section, the first step is to construct battery modules as independent units and then integrate them into a pack to fulfill all requirements and standard regulations. The general trend is towards more compact and lighter systems with reduced costs.
For pouch cells, an energy density of 650~Wh/L and specific energy of 270~Wh/kg can be found in vehicles. For upcoming cells, a 10-20\% enhancement will appear in a couple of years. As an approximation, these numbers can be reduced by approximately 40\% at the pack level. For example, an automotive battery pack using cells with 270~Wh/kg can have an energy density of approximately 180~Wh/kg. 
New cell technologies with improved energy density or specific energy numbers, more integration in packaging, and improved technology for other parts of the system, such as contactors, sensors, and elimination of wires (using wireless), are needed to reduce the weight and volume of the pack. Hence, new materials and chemistry play key roles in meaningful changes in technological advancement.
The battery system voltage is another interesting topic. A nominal voltage of 350~V is widely used for passenger cars. However, in some cases, the system voltage was 700-800~V. The power requirement has increased from less than a hundred kW to more than 200 kW, and in some cases, it has increased even further. Consequently, it is necessary to increase the system voltage level to reduce the current level in the system. Consequently, several high-voltage developments, such as 800~V, are under development by the automotive industry, which will increasingly appear in the market \cite{Haghbin2}. Another point that should be mentioned is the charging infrastructure. 

Another interesting area of research and development in the battery sector involves sodium-based cells, which present an alternative to lithium-ion batteries. Sodium-based batteries utilize sodium ions instead of lithium ions, offering potential advantages such as abundant raw material availability and lower costs, reducing the environmental and ethical concerns associated with lithium and cobalt mining. These sodium-ion batteries are receiving attention for their potential to power EVs in the future, especially as the demand for sustainable and cost-effective solutions. At the begining compared to Li-ion cells, sodium-based cells are expected to play a role in developing batteries which have application in EVs \cite{TNO}.
Another cutting-edge improvements is the development of batteries that do not contain rare earth elements. Many current EV battery chemistries, such as those using LCO battery, depend strongly on rare earth materials. Notwithstanding, as the industry moves toward more sustainable practices, researchers are focusing on new technologies that eliminates or at least minimize the use of these rare earths. These technologies aim to reduce environmental impact and alleviate concerns about supply chain limitations and geopolitical issues surrounding the extraction of these critical materials.
TNO battery technology is characterized by high charging and discharging power capabilities. This innovation provides rapid charge/discharge cycles, which offer the potential for reducing charging times significantly and improving vehicle performance. As the automotive industry continues to focus on reducing charging times and enhancing power delivery, TNO's battery technology could become a key player in meeting these demands \mbox{\cite{TNO_2}}.
Additionally, battery swapping systems, introduced by NIO, offer a unique solution to the issue of long charging times \cite{NIO}. Instead of waiting for a long time to charge battery, NIO offers the ability to swap out the battery entirely in a matter of minutes. This system addresses the challenge of limited fast-charging infrastructure and provides a flexible solution for EV owners, particularly in regions with growing EV adoption but underdeveloped charging networks. Increasing the battery voltage may add issues to the available fast-charging infrastructure. Some fast chargers can not charge 800~V batteries. Hence, there are some works under development to change the classical structure such that the vehicle can utilize the low-voltage charger to charge high-voltage packs.
In addition to passenger cars, battery packs are required for buses, trucks, and aviation vehicles. Usually, the solutions developed for passenger cars are optimized, and the results are expected to be utilized in other sections. 
For aviation electrification, some major hindrances require special attention and further development. The classical structure of a battery pack is inappropriate for aviation. It is not possible to disconnect the battery pack in the case of an extreme fault, and new architectures are required. In addition, the environmental conditions are harsher and packaging may be more challenging. Available aviation solutions, such as satellites, are extremely expensive and not affordable. Hence, more work and maturity are required in this sector.  
%%%%%%%%%%%%%%%%%%%%%%%%%%%%%%%%%%%%%%%%%%%%%%%%%%%%%%%%%%%%
\section{Conclusion}
\noindent The automotive battery pack status, from an automaker's point of view, is reviewed, and near-future expectations and technological advancements are presented. Different parts of the system, including major regulatory standards, are presented, and near-future development activities are mentioned. Pack technology is advancing at a rapid pace, and it is expected to see more effective, safer, and cheaper units shortly. The following points highlight some of the most important ongoing changes and emerging technologies that will drive the next generation of automotive battery systems:
\begin{itemize}
    \item Battery technology is developing and advancements in energy density, safety, and cost-effectiveness are expected shortly.
    \item High voltage systems (700–800~V) are developing due to their high efficiency, reduced power losses, and faster charging. Despite this, they require upgrades in infrastructure and power electronics.
    \item Sodium-based batteries are developing as a great alternative to Li-ion batteries, offering advantages such as abundant raw materials, lower costs, and reduced environmental impact, while they remain in early development.
    \item Rare-earth-free battery chemistries are being explored to reduce reliance on materials like cobalt and lithium, addressing supply chain concerns and sustainability challenges.
    \item TNO battery technology offers high charging and discharging power capabilities, which could significantly reduce EV charging times and improve overall the EV performance. 
    \item Battery swapping systems, such as those introduced by NIO, provide a great alternative to fast charging, enabling quick battery swaps and solving infrastructure limitations.
    \item Battery pack integration is becoming more sophisticated, with wireless communication and modular designs enhancing flexibility, maintenance, and manufacturability.
    \item The role of BMS is expanding lifetime, advanced diagnostics, and cybersecurity measures to enhance performance and longevity of the battery pack.
    \item Safety is a critical priority, with stricter regulatory standards governing battery pack design, testing, and operational safety, particularly for thermal runaway prevention and high-voltage handling.
    \item Thermal management is essential for high-performance BMS, with ongoing progress in cooling solutions for both fast charging and high-power applications.
    \item Charging infrastructure must evolve as well as battery technology, because many existing chargers cannot accommodate 800~V systems, requiring new solutions for compatibility with lower-voltage charging networks.  
    \item Second-life applications for EV batteries are expanding, which includes energy storage for the grid and supporting a more sustainable lifecycle, and reducing environmental impact.  
    \item Battery electrification is extending beyond passenger cars, with significant research efforts in heavy-duty transport, such as buses, trucks, and aviation, each of which needs unique design considerations.  
    \item Aviation electrification have new challenges, requiring alternative battery architectures, enhanced environmental durability, and cost-effective solutions that differ from those used in ground vehicles (buses, cars, etc).  
    \item The sustainable development, cost-effective, and high-performance battery solutions will continue to propel advancements in materials, design, and infrastructure, defining the future of electric mobility.  
\end{itemize}

\section{List of abbreviations}
\begin{enumerate}
\item{ASIL} \quad {Automotive Safety Integrity Level.}
\item{BMS} \quad {Battery Management System.}
\item{ECE} \hspace{0.3cm} {Economic Commission for Europe.}
\item{ECU} \hspace{0.3cm} {Electronic Control Unit.}
\item{EMC} \hspace{0.3cm} {Electromagnetic Compatibility.}
\item{EV} \hspace{0.6cm} {Electric Vehicle.}
\item{FEMA} \hspace{0.1cm} {Failure Mode and Effect Analysis.}
\item{HARA} \hspace{0.05cm} {Hazard Analysis and Risk Assessment.}
\item{HVIL} \hspace{0.1cm} {High-Voltage Interlock Loop.}
\item{ISO} \hspace{0.3cm} {International Organization for Standardization.}
\item{LCO} \hspace{0.2cm} {Lithium Cobalt Oxide.}
\item{LFP} \hspace{0.2cm} {Lithium Iron Phosphate.}
\item{LPM} \hspace{0.2cm} {Liters per Minute.}
\item{MSD} \hspace{0.2cm} {Manual Service Disconnect.}
\item{NMC} \hspace{0.2cm} {Lithium Nickel Manganese Cobalt Oxide.}
\item{SG} \hspace{0.5cm} {Safety Goal.}
\item{UN} \hspace{0.4cm} {United Nation.}
\end{enumerate}
\section{Declerations}
\begin{itemize}
  \item Availability of data and material: Not applicable
  \item Competing interests: Not applicable
  \item Funding: Not applicable
  \item Authors' contributions: Saeid Haghbin, Morteza Rezaei Larijani, MohammadReza Zolghadri, and Shahin Hedayati Kia
  \item Acknowledgements: Not applicable
\end{itemize}
% 

%\begin{acknowledgements}
%\end{acknowledgements}



%
%\newpage
%Bib info.
 

%\bf{If you include a photo:}\vspace{-33pt}
%\begin{IEEEbiography}[{\includegraphics[width=1in,height=1.25in,clip,keepaspectratio]{fig1}}]{Michael Shell}
%Use $\backslash${\tt{begin\{IEEEbiography\}}} and then for the 1st argument use $\backslash${\tt{includegraphics}} to declare and link the author photo.
%Use the author name as the 3rd argument followed by the biography text.
%\end{IEEEbiography}

%Use $\backslash${\tt{begin\{IEEEbiographynophoto\}}} and the author name as the argument followed by the %biography text.
\vfill

% etc
%\end{thebibliography}

\end{document}